\documentclass[twocolumn,showpacs,prb]{revtex4}%
\usepackage{epsfig}
\usepackage{caption}
\usepackage{lscape}
\usepackage{bm}
\usepackage{amsfonts}
\usepackage{amsmath}
\usepackage{amssymb}
\usepackage{graphicx}%
\begin{document}
\title{Electric-Field Induced Resonant Spin Polarization in a Two-Dimensional
Electron Gas}
\author{Yun-Juan Bao and Shun-Qing Shen}
\affiliation{Department of Physics, and Center of Theoretical and Computational Physics,
the University of Hong Kong, Hong Kong}
\date{October 14, 2006.}

\begin{abstract}
Electric response of spin polarization in two-dimensional electron gas with
structural inversion asymmetry subjected to a magnetic field was studied by
means of the linear and non-linear theory and numerical simulation with the
disorder effect. It was found by Kubo linear reponse theory that an electric
resonant response of spin polarization occurs when the Fermi surface is
located near the crossing of two Landau levels, which is induced from the
competition between the spin-orbit coupling and Zeeman splitting. The scaling
behavior was investigated with a simplified two-level model by non-linear
method, and the resonant peak value is reciprocally proportional to the
electric field at low temperatures and to temperature for finite electric
fields. Finally numerical simulation illustrated that impurity potential opens
an enegy gap near the resonant point and suppresses the effect gradually with
the increasing strength of disorder. This resonant effect may provide an
efficient way to control spin polarization by an external electric field.

\end{abstract}
\maketitle

\section{Introduction}

Recently electric or non-magnetic generation of spin polarization in
semiconductors has attracted a lot of interests because of its potential
application in spintronic devices of semiconductors
\cite{Prinz98Science,Wolf01Sci}. In this attempt the motion of electron
interacts with its spin via the spin-orbit coupling, which provides a possible
way to control electron spin by electric fields. Several experiments were
resolved to produce spin polarization in quantum wells
\cite{Sih05NP,Stern06PRL,Yang06PRL,GanichevJMMM}, strained semiconductors
\cite{KatoPRL}, and hole-doped hetero-junction \cite{SilovAPL}. In a
two-dimensional electron gas (2DEG) with structural inversion asymmetry, it
was understood that the spin-orbit coupling generates an effective
momentum-dependent field to induce a net bulk spin polarization by electric
fields or currents \cite{Magarill01SSC,Inoue03,Hu04,Ma04PRB}. Another
mechanism has also been proposed to generate spin polarization in a bulk 2DEG
in the presence of in-plane magnetic fields and electric fields
\cite{Halperin0609}. Electric-induced spin accumulation near the boundary of
sample was already observed experimentally in either n- or p-doped
semiconductors as a consequence of spin Hall effect
\cite{Kato04Sci,Wunderlich05PRL,Sinova06SSC}. The ac-field method has also
been applied to induce and detect spin polarization. The spin orientation was
achieved by the excitation of a high-frequency electric field
\cite{TarasenkoPRB}. An ac-field detection of spin resonance was also
discussed extensively \cite{Loss, Matsunami}.

Generally speaking, we may introduce the electric spin susceptibility
$\chi_{E}^{\alpha\beta}$ to describe the response of spin polarization
$S_{\alpha}$ to an external electric field $E_{\beta}$ \cite{note}
\begin{equation}
S_{\alpha}(E)=S_{\alpha}(0)+\chi_{E}^{\alpha\beta}E_{\beta},
\end{equation}
where $S_{\alpha}(0)$ is the spin polarization in the absence of electric
field. Usually the electric spin susceptibility is a tensor, not a vector. The
spin polarization is determined by the symmetry of spin-orbit coupling of the
system. To generate spin polarization efficiently, a large electric spin
susceptibility is expected. In this paper we propose \emph{an electric
resonant response} of spin polarization in 2DEG with Rashba spin-orbit
coupling. First the linear response theory shows that the electric spin
susceptibility becomes divergent when the crossing point of two Landau levels
is accidentally located near the Fermi surface. The additional degeneracy of
two Landau levels is attributed to competition between the spin-orbit coupling
and the Zeeman splitting. Then a simplified two-level model was proposed to
investigate the scaling behaviors of the resonant electric spin
susceptibility. The resonant values of the electric spin susceptibility decay
with either the applied electric field at low temperatures or with the
temperatures in a weak electric field. Finally we take into account the
disorder effect, and apply the truncation approximation to study the syetem
numerically, which is beyond the linear response theory. The dependence of the
divergent behavior on the electric field and temperature was presented for
finite disorder strengths. It is damped with the electric field and
temperature when the energy scale of electric field and temperature is larger
than the impurity potential. The numerical consequence is consistent with the
result of linear response theory.

\section{Model Hamiltonian and Linear Response Theory}

We start with a 2DEG with Rashba coupling confining in a two-dimensional plane
$L_{x}\times L_{y}.$ The model Hamiltonian in the presence of a perpendicular
magnetic field $B$ is given by%
\begin{equation}
H=H_{0}+H_{R},
\end{equation}
where%

\begin{equation}
H_{0}=\frac{1}{2m^{\ast}}\mathbf{\Pi}^{2}-\frac{1}{2}g_{s}\,\mu_{B}%
\,B\,\sigma_{z}\,,
\end{equation}
and the Rashba coupling%

\begin{equation}
H_{R}=\frac{\alpha}{\hbar}(\Pi_{x}\sigma_{y}-\Pi_{y}\sigma_{x}).
\end{equation}
$m^{\ast}$ is the effective mass of the electron, $\mathbf{\Pi=P+}\frac{e}%
{c}\mathbf{A}$ is the kinetic momentum, $g_{s}$ is the Lande-$g$ factor, and
$\mu_{B}$ is the Bohr magneton. We take the Landau gauge $A=y\,B\,\mathbf{\hat
{x}}$ and consider the periodic boundary condition in $\mathbf{\hat{x}}%
$-direction. The Hamiltonian can be solved analytically with the eigenvalues
\begin{equation}
\varepsilon_{n,s}=\hbar\omega_{c}\left(  n+\frac{s}{2}\sqrt{(1-g)^{2}%
+8\,n\,\eta_{R}^{2}}\right)
\end{equation}
where $n$ is a non-negative integer and $s=\pm1$ \cite{Rashba60,ShenResonance}%
. Here the cyclotron frequency $\omega_{c}=eB/m^{\ast}c,$ the magnetic length
$l_{b}=\sqrt{\hbar c/eB},$ the dimensionless g-factor $g=g_{s}m^{\ast}/2m_{e}$
and effective coupling $\eta_{R}=\alpha/l_{b}\hbar\omega_{c}$. The
corresponding eigenvectors are expressed as
\begin{equation}
\left\vert nks\right\rangle =\left(
\begin{array}
[c]{c}%
ic_{ns}^{+}\phi_{nk}\\
-s\,c_{ns}^{-}\phi_{n-1k}%
\end{array}
\right)  ,
\end{equation}
where $\phi_{nk}$ is the eigenstate of the $n$-th Landau level with $k=p_{x}$
the good quantum number because of the periodic boundary condition along the
$x$-direction. Each Landau level has a degeneracy $N_{k}=L_{x}L_{y}/2\pi
l_{b}^{2}$. $c_{ns}^{\pm}$ are the coupling parameters of spin up and down.
$c_{0,+1}^{+}=1$, $c_{0,-1}^{-}=0$, and $c_{ns}^{\pm}=1/\sqrt{1+(u_{n}\mp
s\sqrt{1+u_{n}^{2}})^{2}}$ with $u_{n}=(1-g)/\sqrt{8n}\eta_{R}$ for $n\geq1$
\cite{ShenResonance}. One of the remarkable features of the spectra is
the\ additional crossing of Landau levels, which is generated by the
competition between Rashba coupling and Zeeman splitting such as
$\varepsilon_{n,s=+1}=\varepsilon_{n+1,s=-1}$ if the integer $n$ and the
magnetic field $B$ satisfy
\begin{equation}
\sqrt{(1-g)^{2}+8\,n\,\eta_{R}^{2}}+\sqrt{(1-g)^{2}+8\,\left(  n+1\right)
\,\eta_{R}^{2}}=2. \label{resonance-n}%
\end{equation}
One important factor in this system is the filling factor, \textrm{i.e.}%
\textit{,} the ratio of number of charge carriers to the Landau degenercay
$\nu=N_{e}/N_{k}=2\pi l_{b}^{2}n_{e}$ ($n_{e}$ is the density of charge
carriers). For a specific density, the filling factor is proportional to 1/B,
and for a specific field, it is proportional to the density $n_{e}$.

Now we come to study the electric response of spin polarization when the Fermi
surface is located near the resonant point. We apply an \emph{in-plane} weak
electric field, say, $E_{y}$ along $y$-direction, then the electric spin
susceptibility can be evaluated by means of the Kubo formula in the weak field
limit \cite{Mahan},%
\begin{align}
\chi_{E}^{\alpha y}  &  =\frac{e\,\hbar}{L_{x}L_{y}}\operatorname{Im}%
{\displaystyle\sum\limits_{nn^{\prime}kk^{\prime}ss^{\prime}}}
\frac{\left(  f_{n^{\prime}s^{\prime}}-f_{ns}\right)  }{(\varepsilon
_{ns}-\varepsilon_{n^{\prime}s^{\prime}})(\varepsilon_{ns}-\varepsilon
_{n^{\prime}s^{\prime}}+i\,\tau^{-1})}\tag{Kubo}\\
&  \times\left\langle nks\right\vert S_{\alpha}\left\vert n^{\prime}k^{\prime
}s^{\prime}\right\rangle \left\langle n^{\prime}k^{\prime}s^{\prime
}\right\vert v_{y}\left\vert nks\right\rangle \nonumber
\end{align}
where $v_{y}$ is the velocity in $y$-direction, $f_{ns}$ are the Fermi-Dirac
distribution function and $\tau$ is the lifetime of the system. The
denominator in the formula indicates that the electric spin susceptibility may
become singular when two energy levels are degenerate near the Fermi surface
and for a long lifetime $\tau,\,$\textrm{i.e.}, $\chi_{E}^{\alpha y}$ may
become divergent. For a finite $\tau$ or temperature it is still finite.

\begin{figure}[ptb]
\centerline{\epsfxsize=8cm\epsfbox{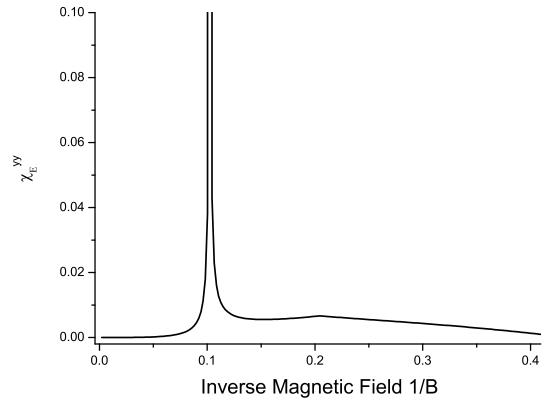}}\caption{$\chi
_{E}^{yy}$ versus $1/B$ at the temperature $T=0$ for weak electric fields by
means of the Kubo formula \ref{Kubo} when $\tau\longrightarrow\infty$. The
parameters are taken as: $\alpha=4.16\times10^{-11}$eV\thinspace m,
$n_{e}=0.118\times10^{-16}$m$^{-2},m^{\ast}=0.05$\thinspace$m_{e},$ $g_{s}=4.$
The unit of $\chi_{E}^{yy}$ is $\hbar/4\pi l_{b}^{2}\,N/C$.}%
\label{PureSystem_SigmaY}%
\end{figure}

For the purpose of numerical calculations, without lossing of generality, we
take the model parameters such that the lowest energy level $\left\vert
1\right\rangle =\left\vert n=0,k,s=+1\right\rangle $ and the first excited
level $\left\vert 2\right\rangle =\left\vert n=1,k,s=-1\right\rangle $ cross
at a critical magnetic field $B=9.8T.$ We plot $\chi_{E}^{yy}$ versus the
inverse of the magnetic field $1/B$ in Fig. \ref{PureSystem_SigmaY} for a
fixed electron density $n_{e}$. It is found that a resonant peak occurs at the
critical magnetic field corresponding to a filling factor $\nu=0.5$. The peak
indicates that it is possible for a weak electric field to induce a finite
spin polarization at the resonant point.

It is worth stressing that the Kubo formula is a result of perturbation and is
valid only when $E\rightarrow0$ (but $E\neq0$)$.$ At a first glance the
divergence of spin susceptibility in Fig. \ref{PureSystem_SigmaY} may be
unphysical because it might be caused by the approaximation of perturbation in
the linear response theory. To clarify the problem, we have to go beyond the
linear response theory and investigate the problem by means of the
non-perturbative approach.

\section{Non-linear Behaviors in Simplified Two-level Model}

\begin{figure}[ptb]
\centerline{\epsfxsize=8cm\epsfbox{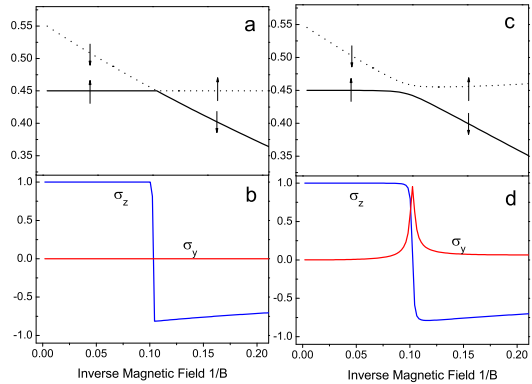}}\caption{(color online)
(a) and (b): Energy spectra for the two degenerate levels in the absence of
the electric field and the correponding spin polarization of the lower-energy
states. (c) and (d): Energies for the two anti-crossing levels in the presence
of the electric field and the corresponding spin polarization for the
lower-energy level. The higher energy states are denoted by dotted line and
the lower are denoted by solid line. The arrows indicate the orientation of
$S_{z}$. The spin polarization $S_{z}$ and $S_{y}$ for the lower-energy states
are denoted by blue and red line respectively. The energy is in unit of
$\hbar\,\omega_{c}$. The parameters are taken the same as in Fig. 1:
$\alpha=4.16\times10^{-11}$eV\thinspace m, $m^{\ast}=0.05$\thinspace$m_{e},$
$g_{s}=4.$}%
\label{2levelCurve}%
\end{figure}

In order to understand the physical origin of the resonance and to analyze the
scaling behaviors of $\chi_{E}^{yy}$ at or near the resonant point, we
consider a simplified two-level model around the resonant point. This is based
on the fact that the dominant contribution to the electric-field-induced
resonance is attributed to the energy crossing of the two levels. This
truncation approaximation reduces an infinite-dimension problem into a
two-level one. As the two-level problem can be solved analytically, the effect
of electric field can be taken into account. The method goes beyond the
perturbation method and the nonlinear behaviors of spin polarization to
external field can be revealed. For this purpose we are only concerned with
the two nearly degenerated levels $\left\vert 1\right\rangle $ and $\left\vert
2\right\rangle $ and ignore the contribution from other levels of higher
energies. Then the total Hamiltonian including the electric potential of
external field can be reduced to a $2\times2$ matrix:%
\begin{equation}
H^{\ast}=\left(
\begin{array}
[c]{cc}%
\varepsilon & v_{E}\\
v_{E} & -\varepsilon
\end{array}
\right)  , \label{effective}%
\end{equation}
where $\varepsilon=\frac{1}{2}\left(  \varepsilon_{1,-1}-\varepsilon
_{0,+1}\right)  $, and $v_{E}=\left\langle 1\right\vert e\,E_{y}\,y\left\vert
2\right\rangle =\eta_{R}\,eEl_{b}c_{0,+1}^{+}c_{1,-1}^{-}$ The off-diagonal
element induced by the electric potential lifts the degeneracy and opens an
energy gap $2\left\vert v_{E}\right\vert $ at the resonant point of
$\varepsilon=0$. In Fig. \ref{2levelCurve}, the energies of the two levels and
the spin polarizations for the lower energy states are plotted with repect to
the inverse of magnetic field. In the absence of electric field the two levels
cross at the resonant point $B=9.8T$ and each state has almost opposite spin
polarization in $z$-direction. From the point of view of the lowest energy
state, the spin polarization $S_{z}$ has a jump near the crossing point as
shown in Fig. \ref{2levelCurve} (b). In the presence of electric field, the
two states of $\left\vert 1\right\rangle $ and $\left\vert 2\right\rangle $
will be admixed due to the off-diagonal term in $H^{\ast}$. The term will lift
the degeneracy of the two levels near the resonant point. In the lowest energy
state, the spin polarization rotates in the $z$-$y$ plane from positive spin
$S_{z}$ to the negative $S_{z}$ when the magnetic field sweeps over the
resonant point. At the resonant point, the system is almost polarized along
the $y$-direction in this two-level problem. Comparing with the case in the
absence of electric field, we notice that non-zero $S_{y}$ is induced by a
weak electric field. Because there is a finite density of states at this point
due to the Landau degeneracy, the total spin polarization in $y$-direction
becomes finite. As a result the electric spin susceptibility $\chi_{E}^{yy}$
becomes divergent at the point.

\begin{figure}[ptb]
\centerline{\epsfxsize=8cm\epsfbox{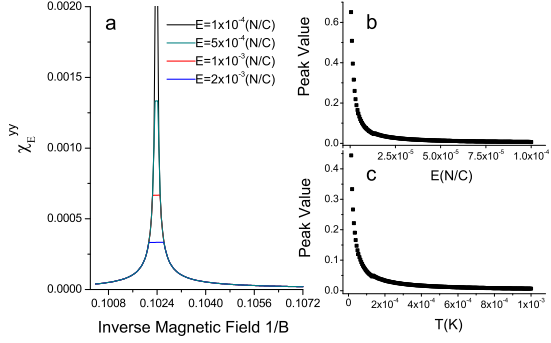}}\caption{(color online)
(a) The spin susceptibility $\chi_{E}^{yy}$ as a function of $1/B$ near the
resonant point for different electric fields at $T=10^{-6}K.$ (b) The electric
field $E$ dependence of the peak value of $\chi_{E}^{yy}$ at a low temperature
$T=10^{-6}K.$ (c) The temperature $T$ dependence of the peak value for an
electric field $E=10^{-6}N/C.$ The unit of $\chi_{E}^{yy}$ is $\hbar/4\pi
l_{b}^{2}\,N/C.$}%
\label{NoDis-Diff-E}%
\end{figure}

If we denote by $S_{y}^{\pm}$ the expectation value of spin polarization
$S_{y}$ of the two split levels, the overall response of the spin polarization
to a finite external field can be calculated by:%
\begin{equation}
\chi_{E}^{yy}=\frac{1}{2\pi l_{b}^{2}}\frac{\left(  S_{y}^{+}\,f_{+}+S_{y}%
^{-}\,f_{-}\right)  _{E}-\left(  S_{y}^{+}\,f_{+}+S_{y}^{-}\,f_{-}\right)
_{E=0}}{E}, \label{2levelSpin}%
\end{equation}
with $f_{\pm}$ the Fermi-Dirac distribution of the two levels and $f_{-}%
+f_{+}=\nu$. With other parameters the same as in Fig.
\cite{PureSystem_SigmaY}, electric spin susceptibility $\chi_{E}^{yy}$ near
the resonant point is plotted as a function of $1/B$ in Fig.
\ref{NoDis-Diff-E} (a) for different electric fields. From Eq.
(\ref{2levelSpin}) we can rewrite it in the form:
\begin{equation}
\chi_{E}^{yy}=\frac{f_{-}(1-f_{+})}{\pi l_{b}^{2}\nu E}\frac{\left\vert
v_{E}\right\vert }{\sqrt{\varepsilon^{2}+\left\vert v_{E}\right\vert ^{2}}%
}\left(  1-e^{-2\sqrt{\varepsilon^{2}+\left\vert v_{E}\right\vert ^{2}}%
/kT}\right)  .
\end{equation}
Near the resonant point $\varepsilon\rightarrow0$, we have $\chi_{E}%
^{yy}\propto1/E$ at low temperatures $kT\ll\left\vert v_{E}\right\vert ,$
meanwhile $\chi_{E}^{yy}\propto1/T$\ for a weak electric field $kT\gg
\left\vert v_{E}\right\vert .$ It depends on the energy scales of the electric
field energy and temperatures. In Fig. \ref{NoDis-Diff-E} (b) and (c) the
dependence of the peak value of $\chi_{E}^{yy}$ on the electric field and
temperature are plotted. A finite spin polarization can be induced by a weak
electric field, but $\chi_{E}^{yy}$ decays with the electric field and the
temperature. This result is consistent with the result of the Kubo formula in
Fig. \ref{PureSystem_SigmaY}. It indicates that the resonance is not caused by
the perturbation approaximation in the linear response approach but is\emph{
attributed to the removal of the degeneracy of two crossing Landau levels by
the external field}\textit{.}

\section{Numerical Simulation}

After establishing a physical picture for the electric resonance of spin
polarization, we come to study the effect of impurities. The effect can be
described by introducing a finite lifetime $\tau$ phenomenally in the Kubo
formula. To go beyond the linear response theory, we do numerical simulation
to study the impurity effect in the real space. For this purpose, we still
take the periodic boundary condition in $x$-direction but an infinite
potential wall in $y$-direction: $V(y)=0$ for $\,\left\vert y\right\vert
<L_{y}/2,\,$and $+\infty$ otherwise. The disorder potential $U(x,y)$ is taken
as the short-range impurities of strength $u_{i}$ uniformly distributed at
$(x_{i},y_{i})$ in the plane \cite{Shizuya}:
\begin{equation}
U(x,y)=%
{\displaystyle\sum\limits_{i}}
u_{i}\,\delta(x-x_{i})\,\delta(y-y_{i})\,,
\end{equation}
where $u_{i}\in(-u/2,u/2),$ $x_{i}\in(-L_{x}/2,L_{x}/2),$ $y_{i}\in
(-L_{y}/2,L_{y}/2).$ In the absence of impurity potential and Rashba coupling,
the confined Landau levels $\left\vert \varphi_{nks}\right\rangle $ have been
obtained analytically \cite{MacDonald1}, which can be regarded as a complete
set of basis. On this basis, the kinetic energy and the Zeeman term, $H_{0}$,
has been diagonalized. The elements of the Rashba coupling and the disorder
potential are $\left\langle \varphi_{n^{\prime}k^{\prime}s^{\prime}%
}\right\vert H_{R}\left\vert \varphi_{nks}\right\rangle $ and $\left\langle
\varphi_{n^{\prime}k^{\prime}s^{\prime}}\right\vert U\left\vert \varphi
_{nks}\right\rangle $, respectively. After taking into account the impurity
potential and Rashba coupling, we apply the truncation approximation to reduce
the whole Hamiltonian into an effective one with a finite dimension.
Furthermore the effective Hamiltonian will be diagonalized numerically to
calculate the eigenvalues and eigenfunctions. In the calculations we take
$n_{\max}+1$ Landau levels $n=0,1,...n_{\max}$ and each Landau level has
$N_{k}$ discrete values of $k,$ ($n=0,1,2,\cdots N_{k}-1$) \cite{Ono3}.
$N_{k}$ can also be explained as the maximum number of electrons accommodated
in each Landau level. Then the number of basis functions we retained in the
truncation approximation is $N=2\times$ $(n_{\max}+1)\times N_{k}$ with double
degeneracy of spin. By diagonalizing the $N\times N$ Hamiltonian numerically,
one get $N$-eigenvalues $E_{m}^{N}$ and $N$-component wave vectors $\left\vert
\Psi_{m}^{N}\right\rangle =%
{\displaystyle\sum}
\alpha_{nks}^{m}\left\vert \varphi_{nks}\right\rangle ,$ which are the
superposition of the basis $\left\vert \varphi_{nks}\right\rangle $. When a
weak electric field $V=e\,E_{y}\,y$ is applied in $y$-direction, the
expectation value of the total spin can be calculated numerically. As a
result, $\chi_{E}^{yy}$ can be obtained%
\begin{equation}
\chi_{E}^{yy}=\frac{1}{L_{x}\,L_{y}}%
{\displaystyle\sum\limits_{m}}
\frac{\delta(S_{y})_{m}}{E_{y}}\, \label{SpinResponse}%
\end{equation}
where
\begin{equation}
\delta(S_{y})_{m}=\left.  f_{m}\left\langle \Psi_{m}^{N}\right\vert
S_{y}\left\vert \Psi_{m}^{N}\right\rangle \right\vert _{E}-\left.
f_{m}\left\langle \Psi_{m}^{N}\right\vert S_{y}\left\vert \Psi_{m}%
^{N}\right\rangle \right\vert _{E=0}.
\end{equation}
with $f_{m}$ the Fermi-Dirac distribution.

In our calculations, the model parameters are taken as $L_{x}/N_{1}%
=L_{y}/N_{2}=\sqrt{2\pi}l_{b}.$ In this paper we take $N_{1}=10,$ and
$N_{2}=6.$ Then the maximum number of electrons at each level is $N_{k}%
=L_{y}/\Delta y_{0}=N_{1}\times N_{2}=60.$ The number of Landau levels are
truncated to $n_{\max}=5,$ hence \thinspace$N=720.$ In this truncation
approximation, we are concerned only with the low energy physics. The magnetic
field is chosen as $B_{c}=9.8T$ such that $E_{0,+1}=E_{1,-1}$ for the lowest
and first excited level in the bulk region, with other parameters the same as
in Fig. \ref{PureSystem_SigmaY}. 140 impurities of relative strength
$\left\vert \lambda_{i}\right\vert =\left\vert u_{i}/(2\pi l_{b}^{2}%
\,\hbar\omega)\right\vert \leq\lambda$ are randomly distributed over the
sample. For each configuration of impurities, both the strength and position
are generated randomly. The states are filled from lower to higher energy, and
correspondingly, $\chi_{E}^{yy}$ can be calculated by the formula in Eq.
(\ref{SpinResponse}) for each configuration.

\begin{figure}[ptb]
\centerline{\epsfxsize=8cm\epsfbox{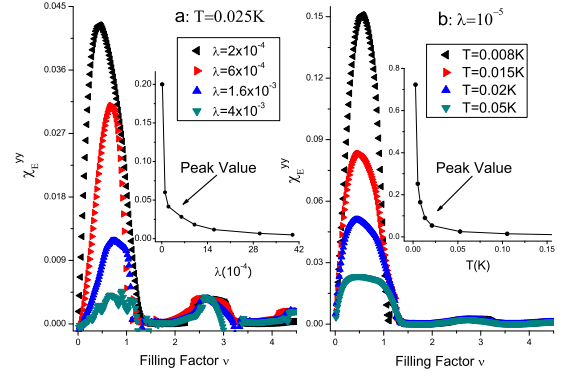}} \caption{(color
online) (a)$\,\chi_{E}^{yy}$ versus filling factor $\nu$ for four different
disorder strength $\lambda$ for a fixed temperature $T=0.025K$. The inset
shows the disorder strength dependence of the peak value. (b) $\chi_{E}^{yy}$
versus the filling factor $\nu$ at different temperatures $T$ for disorder
strength $\lambda=10^{-5}.$ The inset is for the temperature $T$ dependence of
the peak value. The electric field is taken as low as $E_{y}=10^{-8}N/C$ and
the spin susceptibility is in unit of $\hbar/4\pi l_{b}^{2}$\thinspace$N/C.$}%
\label{SigmaY_Vari_Dis_T}%
\end{figure}

\begin{figure}[ptb]
\centerline{\epsfxsize=8cm\epsfbox{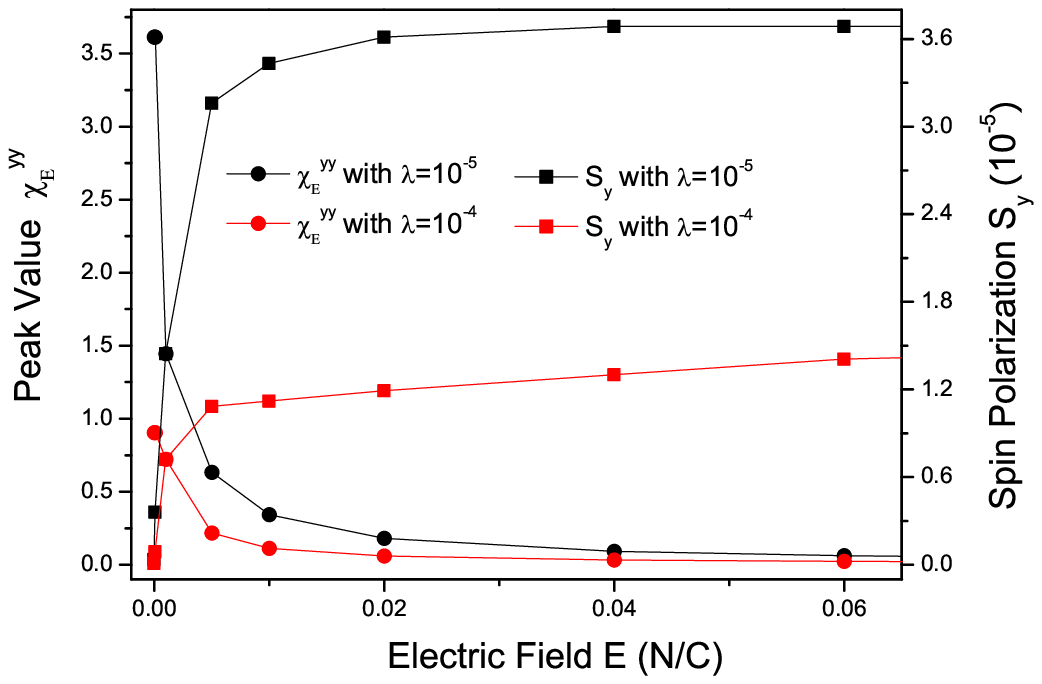}} \caption{(color online)
The electric field $E$ dependence of $\chi_{E}^{yy}$ and the spin density
$S_{y}$ is denoted respectively by the dots and diamonds in the presence of
disorder. The temperature is set to zero. The electric spin susceptibility
$\chi_{E}^{yy}$ is in unit of $\hbar/4\pi l_{b}^{2}\,N/C$ and the spin density
$S_{y}$ is in unit of $\hbar/4\pi l_{b}^{2}.$}%
\label{SigmaY_Vari_E}%
\end{figure}

After averaging over $10^{4}$ different impurity configurations, we plot the
average value of $\chi_{E}^{yy}$ versus the filling factor $\nu$ in Fig.
\ref{SigmaY_Vari_Dis_T} (a) for different strength of disorder $\lambda.$ The
temperature is set to $T=0.025K$ hence the ratio $kT/\hbar\omega_{c}%
=2\times10^{-4}\leq\lambda$ and the electric field $E_{y}=10^{-6}N/C$ such
that $\tilde{e}=eE_{y}l_{b}/\hbar\omega_{c}\simeq10^{-8}<\lambda.$ When the
electrons are filled upto $\nu=0.5,$ $\chi_{E}^{yy}$ displays a resonant peak.
In contrast $\chi_{E}^{yy}$ is finite in the non-degenerate region with the
filling number $\nu=2\sim4$ and tends to be suppressed when $\lambda
=1.6\times10^{-3}$. The relative error due to impurity fluctuation is
estimated to be about $0.01\%$ and upto $5\%$ at the resonant point due to the
fact that the susceptibility is very sensitive to the energy gap opened by
impurities. The peak height decreases with the disorder strength $\lambda$ and
the peak value versus $\lambda$ is demonstrated in the inset. When we
extrapolate to the limit $\lambda\rightarrow\infty,$ the peak value tends to
be suppressed completely. The dependence of the resonant $\chi_{E}^{yy}$ on
the disorder strength is similar to that on the electric field, thus the
impurity scattering opens a gap between the degenerate levels just like the
electric field. Inversely, we fix the disorder strength at $\lambda=10^{-5}$
and plot the electric spin susceptibility $\chi_{E}^{yy}$ in Fig.
\ref{SigmaY_Vari_Dis_T} (b) at different temperatures with $kT/\hbar\omega
_{c}>\lambda.$ The peak height of $\chi_{E}^{yy}$ decays with temperatures.
The scaling behavior of the peak value versus $T$ is shown in the inset. In
the presence of disorder, the dependence of the resonant peak of $\chi
_{E}^{yy}$ on the electric field is also simulated and plotted in Fig.
\ref{SigmaY_Vari_E} denoted by the black and red dot respectively for the
disorder strength $\lambda=10^{-5}$ and $10^{-4}$. At low temperatures, when
$\tilde{e}=eE_{y}l_{b}/\hbar\omega_{c}\ll\lambda,$ $\chi_{E}^{yy}$ is
independent of the electric field $E$ but diverges as $1/E$ when it is
comparable with or greater than the disorder strength $\lambda.$ The spin
density $S_{y}(E)=S_{y}(0)+\chi_{E}^{\alpha\beta}E_{\beta}$ increases linearly
with a weaker electric field, but saturates at higher electric fields as
plotted in Fig. \ref{SigmaY_Vari_E} by the black and red diamond for the
disorder strength $\lambda=10^{-5}$ and $10^{-4},$ respectively.

\section{Summary and Discussion}

At last the occurrence of this resonance is not limited only in the Rashba
system. The electric spin susceptibility depends on the symmetry of spin-orbit
coupling explicitly. The present work can be generalized to a system with the
Dresselhaus coupling, $H_{D}=\beta(p_{x}\sigma_{x}-p_{y}\sigma_{y})$. Because
the Rashba coupling can be transformed to the Dresselhaus coupling under the
transformation of $\sigma_{x}\rightarrow\sigma_{y}$, $\sigma_{y}%
\rightarrow\sigma_{x}$ and $\sigma_{z}\rightarrow-\sigma_{z}$ \cite{Shen04PRB}%
, we conclude that it is $\chi_{E}^{xy}$ instead of $\chi_{E}^{yy}$ which
would become divergent at the resonant point.

In conclusion, a tiny electric field may generate a finite spin polarization
in a disordered Rashba system in the presence of a magnetic field. As a result
the electric spin susceptibility exhibits a resonant peak when the Fermi
surface goes through the crossing point of two Landau levels. Numerical
results demonstrate that the result goes beyond the linear response thoery.
This provides a novel mechanism to control spin polarization efficiently by an
electric field in semiconductors. As the spin polarization can be measured
very accurately it is believed that the effect can be verified in the samples
of 2DEG, such as the hetero-junction of InGaAs/InAlAs.

The work was supported by the Research Grant Council of Hong Kong under Grant
No. HKU 7039/05P and HKU 7042/06P. Numerical work was performed on High
Performance Computing Cluster of Computer Centre of the University of Hong Kong.


\begin{thebibliography}{99}
\bibitem {note}In the linear response theory, the spin susceptibility is
defined as $\chi_{E}^{\alpha\beta}=\left.  \partial S_{\alpha}/\partial
E_{\beta}\right\vert _{E=0}.$ In this paper due to the non-linear behavior of
$\chi_{E}^{\alpha\beta}$ to external field $E$, we use the definition in Eq.
(1) for a finite field.
\end{thebibliography}
\end{document}